\documentclass[openacc]{rstransa}

\usepackage{comment}
\usepackage{grffile}

\newcommand{\av}[1]{{\left\langle #1 \right\rangle}}

\newcommand{\mA}{{\mathcal A}}

\newcommand{\mbR}{{\mathbb R}}

\newcommand{\mRe}{{\mathrm Re}}

\newcommand{\bF}{{\bf F}}

\newcommand{\bO}{{\bf O}}

\newcommand{\bX}{{\bf X}}
\newcommand{\bU}{{\bf U}}
\newcommand{\bV}{{\bf V}}

\newcommand{\bv}{{\bf v}}
\newcommand{\bu}{{\bf u}}
\newcommand{\bx}{{\bf x}}

\newcommand{\eg}{\emph{e.g.}}

\newcommand{\be}{\begin{equation}}
\newcommand{\ee}{\end{equation}}
\newcommand{\bs}{\begin{split}}
\newcommand{\es}{\end{split}}

\newcommand{\hs}{\;\;\;\;\;\;\;\;\;\;\;\;}

\graphicspath{{Logos/}{Figs/v5/}}
\begin{document}

\title{Hidden scale invariance in Navier-Stokes intermittency}

\author{Alexei A. Mailybaev and Simon Thalabard}

\address{IMPA, Rio de Janeiro, Brazil}

\subject{fluid mechanics, differential equations, mathematical physics}


\corres{Alexei A. Mailybaev\\
\email{alexei@impa.br}\\
Simon Thalabard\\
\email{simon.thalabard@ens-lyon.org}}

\begin{abstract}
We expose a hidden scaling symmetry  of the Navier-Stokes equations   in the limit of vanishing viscosity,  which stems from dynamical space-time rescaling around  suitably defined Lagrangian scaling centers. 
At a dynamical level, the hidden symmetry projects solutions which differ up to  Galilean invariance and global temporal scaling onto the same representative  flow.  At a statistical level, this projection repairs the scale invariance, which is broken by intermittency in the original formulation. Following previous work by the first author,  we here postulate and substantiate with numerics that hidden symmetry statistically holds in the inertial interval of fully developed turbulence. We show that this symmetry accounts for the scale-invariance of a certain class of observables, in particular, the Kolmogorov multipliers.
\end{abstract}


\begin{fmtext}
\section{Introduction}
The notion of statistical  symmetries shapes the modern description of fully developed turbulent states, along with the physical postulate that most of the symmetries of the Navier-Stokes (NS) equations,
describing the dynamics of an incompressible  velocity field $\bu(\bx,t)$ in three-dimensional space and  
 broken by the presence of driving and dissipative mechanisms, 
  are restored in a statistical sense in a suitable limit of small scales $\ell$ and large Reynolds numbers \cite{frisch1995turbulence}.
The postulate is apparently contradicted  by observations.
The global symmetries of the NS system feature  space-time translations, parity, rotations, Galilean transforms and space-time scalings  with suitable rescaling of forcing $\mathbf{f}$ and viscosity $\nu$; see Table~\ref{table:symmetries}.
While the NS equations indeed produce solutions with plausibly isotropic and homogeneous statistics at small scales, these solutions are intermittent. A measurable effect is the power-law behavior of  structure functions $S_p(\ell) = \av{\delta u_\ell^p } \propto \ell^{\zeta_p}$ for longitudinal velocity fluctuations \cite{frisch1995turbulence}, where the exponents $\zeta_p$ depend nonlinearly on the order $p$. This implies breaking of the statistical scale-invariance, \emph{i.e.} the shapes of the  distributions depend on the observation scale $\ell$; see Fig.~\ref{fig:non-Gaussian}.
\end{fmtext}

\maketitle
\noindent 
 
 \begin{table}[t]
\caption{Symmetries of the  NS equations  considered in $\mbR^3$. 
}
\label{table:symmetries}
\begin{tabular}{lllllll}
\hline\hline
\\[-8pt]
 	 & parameters\ \  &$t\mapsto \hs$  &$\bx \mapsto \hs$ &$\bu \mapsto  \hs$ & $\nu \mapsto  \hs $& $\mathbf{f} \mapsto  $ 
\\[3pt]
\hline
\\[-8pt]
Galilean & $\bu_0 \in \mbR^3$ &$t$  &$ \bx+t \bu_0 $ &$ \bu+\bu_0$ & $ \nu$& $  \mathbf{f}$ 
\\[3pt]
Rotation &$\bO\in \mathrm{SO}(3) $ &$ t$  &$ \bO\bx$ &$ \bO\bu$ & $\nu$& $  \bO \mathbf{f}$ 
\\[3pt]
Parity & &$ t$  &$ -\bx$ &$ -\bu$ & $   \nu$& $  - \mathbf{f}$ 
\\[3pt]
Translation & $\Delta t \in \mbR, \; \Delta \bx \in \mbR^3$ &$ t+\Delta t$  &$ \bx+\Delta \bx$ &$  \bu$ & $   \nu$& $   \mathbf{f}$
\\[3pt]
Scaling & $\alpha,\lambda >0 $&$ \alpha t $ & $ \lambda \bx$& $ (\lambda/\alpha) \bu$  &  $ (\lambda^2/\alpha)\nu$   & $ (\lambda/\alpha^2)\mathbf{f}$   
\\[3pt]
\hline\hline
\end{tabular}
\vspace*{-4pt}
\end{table}

\begin{figure}[t]
\centering\includegraphics[width=0.49\textwidth]{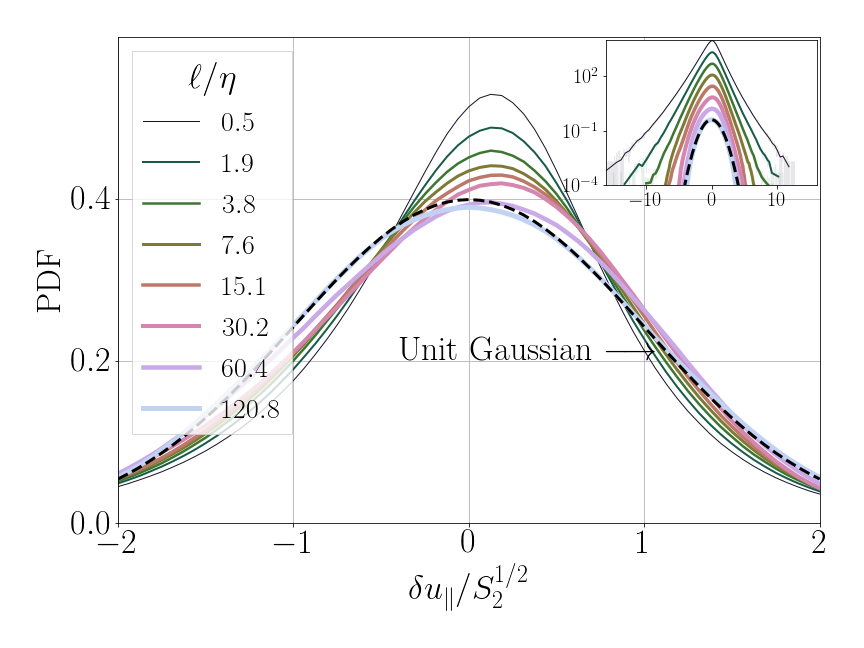}
\centering\includegraphics[width=0.49\textwidth]{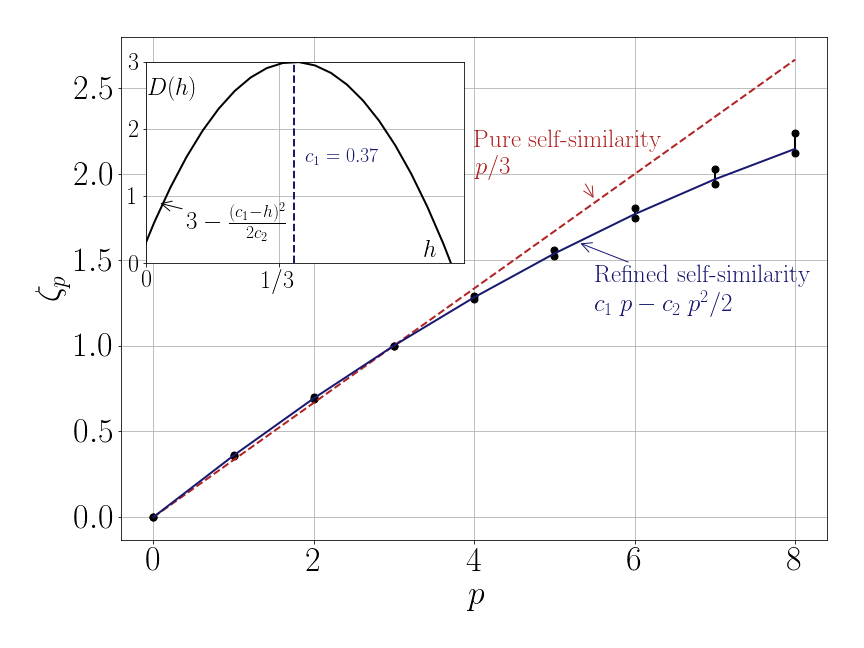}
\caption{{\bf Textbook vision of intermittency.} {\bf Left:} Normalized distributions of parallel velocity increments in $512^3$ DNS at $Re\sim 2600$ with $L/\eta \sim 60$. The inset shows  the same data using semi-log representation with data  vertically shifted for clarity, displaying evolution from nearly Gaussian shaped to heavy tailed distributions as $\ell$ decreases from large to small scales.  {\bf Right: } Scaling exponents  of structure functions, along with the refined self-similarity  quadratic fit, with $c_2= 0.026$ and $c_1 =3 c_2/2+1/3\simeq 0.37$. The inset shows the corresponding singularity spectrum centered around $c_1$, obtained as the Legendre transform $3-\min_p\left\lbrace\zeta_p-ph\right\rbrace$ \cite{frisch1985singularity}; see Table \ref{table:dns}
in \S \ref{sec:five} for physical parameters and \cite{allende2021dynamics} for details on the numerics.}
\label{fig:non-Gaussian}
\end{figure}

\noindent\textbf{Disentangling the scaling symmetries.}
One could argue, however, that intermittency reflects a degeneracy rather than breaking of scaling symmetries, meaning that turbulent limit restores simultaneously all of the allowed scaling symmetries in an entangled way.
One fruitful way to reveal the presence of these symmetries is to invoke the Parisi-Frisch multifractal framework and decompose the  turbulent flow into a superposition of  fractal sets, each set supporting the space-time scaling symmetry 
	\begin{equation}
	\label{eqI_1}
    t,\bx,\bu \to  \lambda^{1-h}t, \lambda  \bx, \lambda^h \bu
    \end{equation}
for some $h \in \mbR $. These sets have dimensions $D(h)$, which define the so-called ``spectrum of singularities'' related to the  exponents $\zeta_p$ by the Legendre transform \cite{frisch1985singularity,argoul1989wavelet,jaffard2000frisch,chevillard2006unified}. In fully developed turbulence, $D(h)$ is a bell shaped function centered slightly above $1/3$, which implies the presence of entangled scaling symmetries; see Fig. \ref{fig:non-Gaussian}.

The  fact that the local structure of turbulent fluctuations entails  local rather than global form of scaling is an idea  at the core of many  heuristic models of intermittency. The lognormal model imagined  by Kolmogorov \cite{kolmogorov1962refinement} relies on a constitutive tie between the local velocity field and the locally averaged dissipation field. Similar models \cite{she1994universal, dubrulle1994intermittency, ruelle2017theory} prove remarkably efficient at describing the nonlinear behavior of $\zeta_p$'s with little or no adjustable parameters, and it is tempting to believe that their hierarchical phenomenology reflects some type of \emph{hidden symmetry} of the NS dynamics, as suggested in  \cite{she1994universal,dubrulle1994intermittency}. 
Such connection is however lacking, and the purpose of our paper is to suggest a candidate symmetry allowing to fill this gap.

\begin{figure}[t]
\centering\includegraphics[width=0.49\textwidth]{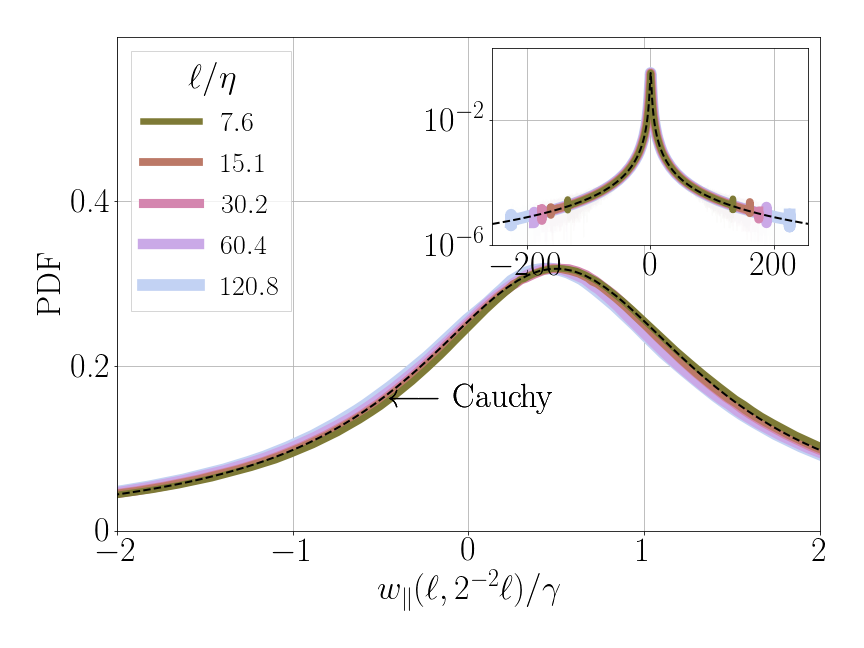}
\centering\includegraphics[width=0.49\textwidth]{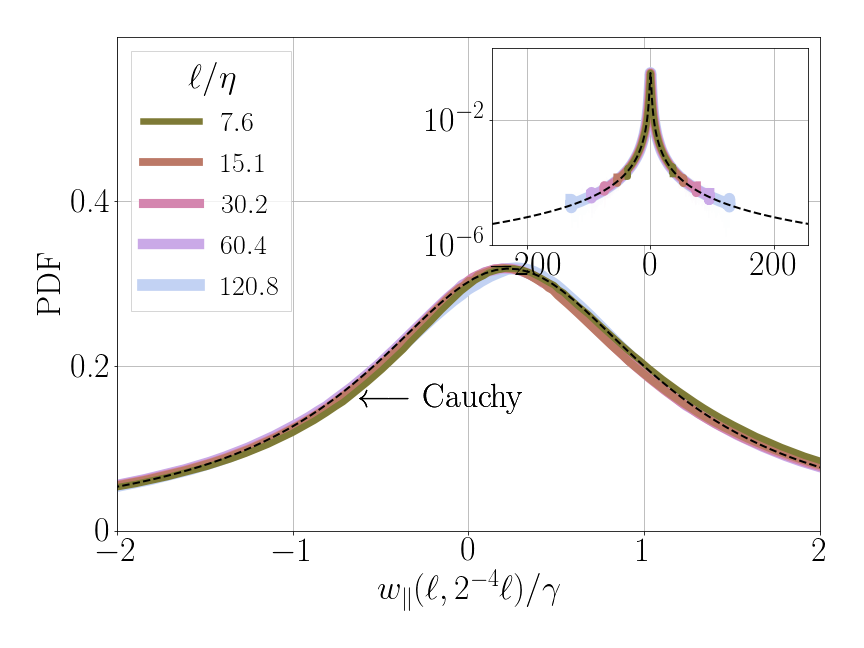}
\caption{{\bf Plausibility of the intrinsic refined self-similarity.} Normalized distributions of multipliers $w_\parallel(\ell,\ell'):=w_{xx,x}(\bx,t;\ell, \ell')$ for the  $512^3$ DNS of Fig.~\ref{fig:non-Gaussian} at $Re\sim 2600$ with $L/\eta \sim 60$, for ratio $\ell'/\ell = 2^{-2}$ (left) and $\ell'/\ell = 2^{-4}$ (right). The  scale factor is $\gamma=1/(\pi  p_{\max})$ with $p_{\max}$ the PDF maximum. The insets  show the tails of the distributions using semilog representations. The dashed black lines show the Cauchy distributions {$p(w)=\frac{1}{\pi} \frac{1}{(w-\bar w_\lambda)^2+1}$}~centered around $\bar w_{1/4} = 0.50$ (left) and $\bar w_{1/16} = 0.23$ (right).}
\label{fig:multipliers}
\end{figure}

\vspace{2mm}\noindent\textbf{Kolmogorov multipliers and hidden scaling symmetry.}
\newcommand{\rrss}{RSH}
Our focus here is the intrinsic version of the refined self-similarity suggested by Kolmogorov in the last three paragraphs of his 1962 paper~\cite{kolmogorov1962refinement}. The original refined similarity hypotheses are there  freed from the extrinsic choice of dissipation statistics by assuming the universality of so-called \emph{Kolmogorov multipliers}. These multipliers are ratios of velocity increments at two different scales $\ell_1$ and $\ell_2$ defined as  
\be
    \label{eq:turbulentmultipliers}
	w_{ij,k}(\bx,t;\ell_1, \ell_2)
	:= \dfrac{u_k(\bx+\ell_1\mathbf{e}_i,t)-u_k(\bx,t)}{u_k(\bx+\ell_2\mathbf{e}_j,t)-u_k(\bx,t)},
\ee
where $\mathbf{e}_i$ is the unit vector in $i$th direction.
 Statistics of these quantities prove to be remarkably self-similar, depending only on the ratio $\ell_2/\ell_1$ and not on the scales themselves. The universality holds even at moderate Reynolds number, with the resulting distributions convincingly  approximated by Cauchy distributions; this was  first noticed  in \cite{chen2003kolmogorov} and is here illustrated in Fig.\ref{fig:multipliers}. 
  As noticed by \cite{chen2003kolmogorov},  the algebraic tails of the multiplier statistics are deceptive: They originate from vanishing denominators in Eq.~(\ref{eq:turbulentmultipliers}) and, as a consequence,  structure functions diverge in the absence of correlations among multipliers. However, the universality  of multipliers statistics has a broader theoretical significance, revealing scale invariance in intermittent dynamics and justifying turbulence models based on random multiplicative cascades \cite{frisch1995turbulence,ruelle2017theory}.
 Although the intrinsic refined self-similarity has gained renewed attention in the context of shell models \cite{benzi1993intermittency,eyink2003gibbsian,mailybaev2020hiddenb,vladimirova2021fibonacci}, it has comparatively been little studied in the turbulent literature.

The purpose of our paper is to connect the intrinsic refined self-similarity  to the general concept of hidden scaling symmetry recently formulated in \cite{mailybaev2020hiddena}. For the NS equations, hidden symmetries are symmetries of suitably transformed solutions, where both space and time are dynamically rescaled in a referential frame defined along a Lagrangian trajectory.
This dynamical rescaling leads to a collapse of the scaling symmetries 	(\ref{eqI_1}) with different exponents $h$ into a single symmetry, therefore removing the degeneracy of scaling symmetries, and providing the mechanism to reveal scale-invariant nature of intermittency. This argument extends the intrinsic self-similarity hypothesis to a larger class of observables, expressed as velocity increments suitably rescaled by positive definite quantities.

The paper is organized as follows. \S \ref{sec:two}  introduces the dynamical rescaling of the NS equations, and  \S \ref{sec:three}  introduces the notion of \emph{hidden symmetries}, namely the symmetries of the rescaled system.  \S\ref{sec:four} relates the hidden symmetries to the scaling symmetries of the original NS system, and points out that hidden scale invariance represents a  weaker type of scale invariance,  which might very well be restored in spite of the breaking  of usual scale invariance. \S\ref{sec:five} shows that  the statistical hidden scaling symmetry implies in particular the scale-invariance of Kolmogorov's multipliers. 
\S\ref{sec:six} describes statistical analysis from large direct numerical simulations beyond the case of multipliers, illustrating possible validity of hidden scale invariance for inertial range statistics. \S\ref{sec:seven} formulates concluding remarks. The Appendix contains technical derivations.

\section{Dynamical rescaling of the Navier--Stokes system}
\label{sec:two}
We consider the dimensionless incompressible Navier--Stokes system 
	\begin{equation}
	\tag{NS}
	\label{NS}
	\frac{\partial\bu}{\partial t} + \bu\cdot\nabla \bu 
	+ \nabla p= \mRe^{-1}{\Delta\bu } +\mathbf{f}, \quad
	\nabla\cdot\mathbf{u} = 0,
	\end{equation}
for the velocity field $\bu(\bx,t)$, the Reynolds number $\mRe = UL/\nu$ and the force $\mathbf{f}(\bx,t)$. 
 For simplicity, we assume that the velocity and the force are periodic in space $\mbR^3$ with no physical boundary, and that forcing is solenoidal (divergence-free).  
We now describe the dynamical space-time rescaling of the NS system, which is achieved in three steps.

\paragraph{The first step} \hspace{-0.3cm} consists in defining the scaled relative velocity field as
\be\label{eq_resc_sp}
	\Delta\bu_\ell(\bx,\bX,t) :=  \bu\big(\bx+\ell\bX,t\big)-\bu\big(\bx,t\big),
\ee
where $\bx$ is a position of a new reference frame, $\bX$ is a rescaled coordinate, and $\ell > 0$ is a given ``zooming'' scale. At each position $\bx$ and time $t$, we introduce the local velocity amplitude $A_{\ell}(\bx,t)$ and time scale $T_{\ell}(\bx,t)$ as
\be
    \label{eq2_4}
   A_{\ell}(\bx,t) := \mathcal{A}[\Delta \mathbf{u}_\ell(\bx,\cdot,t)],\quad 
   T_{\ell}(\bx,t) := \frac{\ell}{A_{\ell}(\bx,t)},
\ee
where $\mathcal{A}[\bV]$ is a prescribed functional acting on vector fields $\bV(\bX)$ and defining a local average with the homogeneity property
	\begin{equation}
	\label{eq_A_hom}
    \mathcal{A}[c\mathbf{V}] = c\mathcal{A}[\mathbf{V}] \quad 
    \textrm{for} \quad c > 0. 
    \end{equation}
This functional must be positive for velocity fields under consideration.
For all practical purposes, the reader may think of this functional in terms of a local root-mean-squared velocity, \eg
\be \label{eq_cases}
	\mathcal{A}[\mathbf{V}] := 
	\sqrt{\langle\|\mathbf{V}\|^2 \rangle_{B}},
\ee
where $\langle \cdot \rangle_B$ denotes the average in the unit ball $\|\bX\| \le 1$. The transformations and subsequent algebraic manipulations below  are however fully general, and formally valid  for any positive functional satisfying  the homogeneity property~(\ref{eq_A_hom}); this freedom of choice will be further discussed in  the numerical tests of \S~\ref{sec:six}.

\paragraph{The second step} \hspace{-0.3cm} consists in performing  time rescaling of the field (\ref{eq_resc_sp}) in 
a \emph{quasi-Lagrangian} reference frame.
Thereby, we consider the \emph{scaling center} $\bx_*(t)$ following the Lagrangian  trajectory of an arbitrary fluid parcel (tracer)
and prescribed by the equations
	\begin{equation}
	\label{eq:Xstar}
	\frac{d\mathbf{x}_*}{dt} = \bu \left(\bx_*,t\right)
	, \quad \bx_*(0) = \bx_0,
	\end{equation}
for an arbitrary initial point $\bx_0$. 
Then, by setting $\bx = \bx_*(t)$ and using local scales (\ref{eq2_4}), we define the proper velocity field %
$\bU(\bX,\tau;\bx_0,\ell)$
depending on the proper time variable $\tau$ as
	\begin{equation}
	\label{eq:newvariables}
	\bU(\bX,\tau;\bx_0,\ell)
	:= \frac{\Delta \mathbf{u}_\ell(\bx_*(t),\bX,t)}{A_\ell(\bx_*(t),t)},\quad 
	\tau := \int_0^{t}\frac{ds}{T_\ell(\bx_*(s),s)}.
	\end{equation}	
This transformation depends on the chosen trajectory via its initial point $\bx_0$ and on the scale $\ell$;
In the remainder of the paper, we do not designate this dependence explicitly in our notation of the field $\mathbf{U}(\bX,\tau)$.
Expressions (\ref{eq_resc_sp}) and (\ref{eq:newvariables}) represent two simultaneous local scalings: the spatial scaling with factor $\ell$ around the scaling center $\bx = \bx_*(t)$ and a time-dependent temporal scaling with the factor $T_\ell(\bx,t)$. These scalings are tuned such that the resulting field has typical variations $\|\mathbf{U}\| \sim 1$ at distances $\|\bX\| \sim 1$ and times $\delta \tau \sim 1$. 
One can check that 
	\begin{equation}
	\label{eq2_star}
    \bU(\mathbf{0},\tau) = 0, \quad \mathcal{A}[\bU(\cdot,\tau)] = 1,    \end{equation}	
at any time $\tau$ by construction.
The rationale beneath the proposed space-time scaling  relates to symmetries or projections thereof; we defer the discussion to  \S~\ref{sec:four} \& \ref{sec:five}.

\begin{figure}[t]
\centering\includegraphics[width=0.8\textwidth]{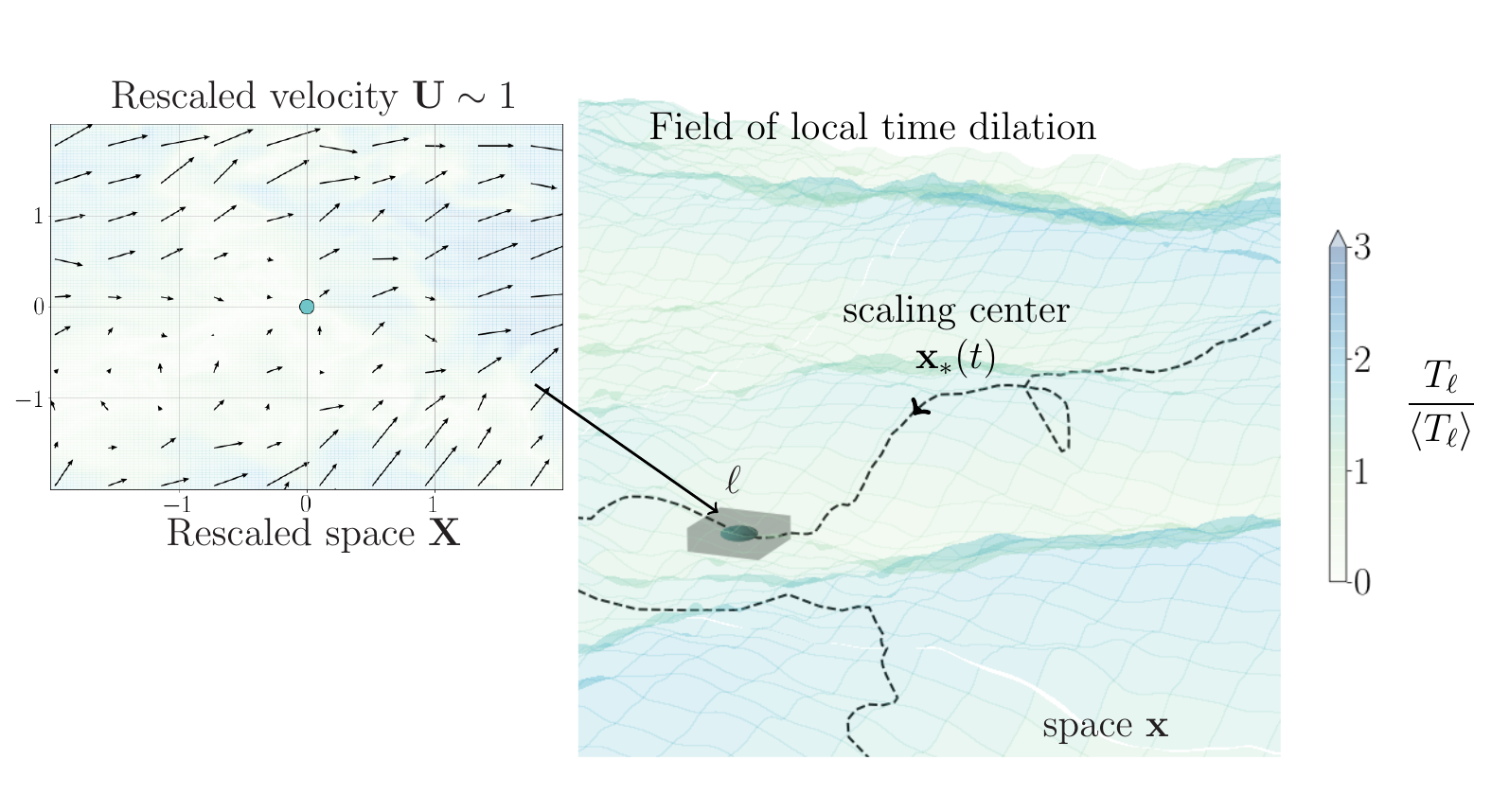}
\caption{Impressionistic view of a turbulent landscape, as seen from the scaling center perspective. In the main panel, the color intensity encodes the turbulent activity, measured through the local time dilatation factor $T_\ell(\bx,t)/\langle T_\ell\rangle$: the proper time $\tau$ flows faster in the lighter region than in the darker ones.
 The left inset shows the rescaled velocity field $\bU(\bX,\tau)$ corresponding to the small gray region.}
\label{fig:rescaling}
\end{figure}

Heuristically, our transformation describes the flow in a reference frame moving with a fluid parcel and evolving in a proper time $\tau$, whose change is synchronized with the flow activity at scale $\ell$. Thus, $\tau$ changes slower when local velocity fluctuations are small, accelerating when the Lagrangian particle $\bx_*(t)$ enters more active regions; see Fig. \ref{fig:rescaling}.
In the usual language of turbulent flows, this  ``time dilatation''  is connected to fluctuations of local eddy turnover time $T_\ell(\bx,t)$ around its space-time average value $\langle T_\ell\rangle$ with the dilation factor $T_\ell(\bx,t)/\langle T_\ell\rangle$.

\paragraph{The third and final step} \hspace{-0.3cm} consists in deriving the dynamics in terms of the new field. Performing the change of variables
$t, \bx, \bu \to  \tau, \bX, \bU$ in the  NS system, and following  the calculations described in Appendix \ref{app:derivation}, one obtains the new equations
\be
	\label{eq:newsystem}
	  \partial_\tau \bU = 
	  \Lambda_\bU \left[-\bU \cdot \nabla \bU
	  -\nabla P
	  +\mRe_{*\ell}^{-1} \Delta \bU +\bF_{*\ell}\right],
	  \quad \nabla\cdot \bU = 0,
\end{equation}
which govern the dynamics of field $\bU(\bX,\tau)$.
Here the operators $\nabla$ and $\Delta$ apply to the coordinates $\bX$.
The linear operator $\Lambda_\bU$ acting on a vector field $\mathbf{V}(\bX)$ at a given time $\tau$ is defined as 
\be
	\label{eq:LambdaU}
	   \Lambda_\bU[\bV] =  \widetilde{\bV}
	   -\bU(\cdot,\tau) \left(
	   \left.\frac{\delta\mathcal{A}}{\delta\bV}\right|_{\bU(\cdot,\tau)}\widetilde{\bV}\right)
\ee 
with $\widetilde{\bV}(\bX) = \bV(\bX)-\bV(\textbf{0})$. We here emphasize that this variational expression is valid for an arbitrary choice of $\mathcal{A}[\cdot]$ with property (\ref{eq_A_hom}). In the case (\ref{eq_cases}) we can compute the variational derivative in (\ref{eq:LambdaU}) explicitly. Since $\mathcal{A}[\bU] = 1$, this yields $\Lambda_\bU[\bV] = \widetilde{\bV} - \bU\big\langle\bU \cdot \widetilde{\bV}\big\rangle_B$. As explained in Appendix \ref{app:derivation}, the new pressure $P(\bX,\tau)$ is defined by the Poisson equation $\Delta P = - \text{Tr} (\nabla \bU)^2$ following from the incompressibility condition, just as for the old variables. The time-dependent Reynolds number and forcing term are defined as
\be
	\label{eq:newsystem_terms}
    \mRe_{*\ell}(\tau) :=  \ell A_\ell(\bx_*(t),t) \,\mRe,
    \quad
    \mathbf{F}_{*\ell}(\bX,\tau)
    :=\dfrac{\ell \mathbf{f}\left(\bx_*(t)+\ell\bX,t\right)}{A^2_\ell(\bx_*(t),t)},
\end{equation}
with $\tau$ given in (\ref{eq:newvariables}); 
these two quantities depend on the choice of the Lagrangian trajectory $\bx_*(t)$ and on the scale $\ell$.

\section{Hidden symmetries of the rescaled dynamics}
\label{sec:three}
We now wish to consider the proper fields $\bU(\bX,\tau)$ obtained for scales $\ell$ lying within the inertial interval of fully developed turbulence. Mathematically speaking, this corresponds to the double limit $\mRe \to \infty$ and $\ell \to 0$ taken in that order. 
We show that the rescaled dynamics features several symmetries, involving either linear or nonlinear transforms of the proper field. 
The linear  symmetries relate to rotation, parity and proper time translation, analogous to those for the old (non-rescaled) velocity field.
The nonlinear symmetries relate to scaling and space translation. Due to their intricate properties which will further be discussed in \S\ref{sec:four}, we interpret them as \emph{hidden symmetries} of the original dynamics.

\paragraph{Rescaled system in the inertial range.}
First, we verify that both dissipative and forcing terms in the rescaled  dynamics~(\ref{eq:newsystem}) become negligible in the inertial interval. 
By construction, the field $\bU(\bX,\tau)$ describes fluctuations of order $\|\bU\| \sim 1$ at distances $\|\bX\| \sim 1$ with time intervals $\delta\tau \sim 1$. Consider now the forcing term $\mathbf{F}_{*\ell}$ defined in  (\ref{eq:newsystem_terms}). In the dimensionless formulation of the NS system, we have $\|\mathbf{f}\| \sim 1$. The Kolmogorov (K41) estimate for velocity fluctuations yields $A_\ell(\bx,t) \sim \ell^{1/3}$, with the intermittency introducing a small correction in the exponent \cite{frisch1995turbulence}. Hence, the forcing term $\mathbf{F}_{*\ell}$ becomes negligible for 
 	\begin{equation}
	\label{eq3_1}
	\ell  \ll 1.
	\end{equation}
This condition designates the distancing from the forcing range. Similarly, a magnitude of the dissipative term is given by $\mRe_{*\ell}^{-1}$ defined in  (\ref{eq:newsystem_terms}). Hence, the dissipative term is negligible for
 	\begin{equation}
	\label{eq:inertial1}
	\ell  \gg \eta_\ell(\bx_*(t),t), \quad \eta_\ell(\bx,t) = \frac{1}{A_\ell(\bx,t)\mRe},
	\end{equation}
where $\eta_\ell(\bx,t)$ denotes the field of local viscous scales. In spite of $\eta_\ell(\bx,t)$ being an intermittently fluctuating quantity, it is expected to vanish in the limit of large Reynolds numbers. In particular, using the Kolmogorov (K41) estimate $A_\ell(\bx,t) \sim \ell^{1/3}$ in (\ref{eq:inertial1}) yields the well-known condition $\ell \gg \eta_K$ with Kolmogorov's   scale $\eta_K = \mRe^{-3/4}$.  

Under conditions (\ref{eq3_1}) and (\ref{eq:inertial1}), or in plain words for the scales $\ell$ in the inertial interval, we can neglect both the forcing and dissipative terms in system (\ref{eq:newsystem}). This  yields 
\be
	\label{eq:newsystem_inviscid}
		 \partial_\tau \bU 
	 = \Lambda_\bU [-\bU \cdot \nabla \bU-\nabla P],
    \quad
	 \nabla\cdot \bU = 0,
\ee
with the pressure defined by the Poisson equation $\Delta P = - \text{Tr} (\nabla \bU)^2$.
This system formally corresponds  to the rescaled Euler system for ideal incompressible flow. 

\paragraph{Explicit linear symmetries: rotation, parity, time-translation.}
Similar to the original system, the rescaled dynamics features explicit, linear transforms as symmetries.
Those symmetries are summarized in the first three lines of Table~\ref{table:hiddensymmetries}, and relate to rotation, parity, and time-translation symmetries.
Their validity can be checked either directly from manipulating the inviscid new system (\ref{eq:newsystem_inviscid}), or by first transforming the old field with respect to any of the symmetries  and then performing the dynamical rescaling.
We note that the rotation and parity symmetries require  the functional $\mathcal{A}[\cdot]$ to be  invariant under those symmetries, {\it i.e.}, \;$\mA[\bV]=\mA[\bV']$ for $\bV'(\bX)=-\bV(-\bX)$ and $\bV'(\bX)=\bO\bV(\bO^{-1}\bX)$, $\bO \in \mathrm{SO}(3)$.

\paragraph{Hidden scaling symmetry.} To reveal the first nonlinear symmetry, we make the obvious but crucial observation that the scale $\ell$ does not explicitly appear in system (\ref{eq:newsystem_inviscid}). The change of scale $\ell \to \ell'$ induces a symmetry transformation $\bU (\bX,\tau) \mapsto \bU' (\bX,\tau')$ for solutions of equation (\ref{eq:newsystem_inviscid}), which we  explicitly compute using definitions (\ref{eq_resc_sp}), (\ref{eq:newvariables}) and property (\ref{eq_A_hom}). Calculations of Appendix~\ref{sec:hiddenscaleinvariance} express the field $\bU'(\bX,\tau')$ corresponding to $\ell'$ in terms of the field $\bU(\bX,\tau)$ corresponding to $\ell$ as 
	\be
	\label{eq3_4}
    \ell \mapsto \ell':\quad \bU' (\bX,\tau') 
    = \frac{\bU_\lambda(\bX,\tau)}{\mathcal{A}
    [\bU_\lambda(\cdot,\tau)]},
	\ee
where 
	\be
	\label{eq3_5}
    \tau' = \lambda \int_0^{\tau}\mathcal{A}
    [\bU_\lambda(\cdot,s)]\,d s,\quad 
    \bU_\lambda(\bX,\tau) := \bU\left(\frac{\bX}{\lambda},\tau\right),\quad 
    \lambda = \frac{\ell}{\ell'}.
\ee
The transformation (\ref{eq3_4})--(\ref{eq3_5}) depends only on the scaling parameter $\lambda = \ell/\ell'$ and maps solutions $\bU(\bX,\tau)$ to new solutions $\bU'(\bX,\tau')$ of system (\ref{eq:newsystem_inviscid}). Therefore, it represents a  space-time symmetry of this system, which is nonlinear and nonlocal. We call it the \textit{hidden scaling symmetry}.

\paragraph{Hidden translation symmetry.}
Similarly, we observe that the choice of the Lagrangian scaling center $\bx_*(t)$, which is defined by the initial point $\bx_0$, does not explicitly appear in System (\ref{eq:newsystem_inviscid}). Consider two different initial points $\bx_0$ and $\widetilde{\bx}_0$ of the Lagrangian trajectories $\bx_*(t)$ and $\widetilde{\bx}_*(t)$. The corresponding fields $\bU (\bX,\tau)$ and $\widetilde{\bU} (\bX,\widetilde{\tau})$ are related as (see Appendix~\ref{sec:hiddenscaleinvariance})  
	\be
	\label{eq3_6}
    \bx_0 \mapsto \widetilde{\bx}_0:\quad \widetilde{\bU} (\bX,\widetilde{\tau}) 
    = \frac{\Delta \bU(\bX_*(\tau),\bX,\tau)}{\mA
    \left[\Delta \bU\left(\bX_*(\tau),\cdot,\tau\right)\right]},
	\ee
where 
	\be
	\label{eq3_7}
    \widetilde{\tau} = \int_0^{\tau}\mathcal{A}
    \left[\Delta \bU(\bX_*(s),\cdot,s)\right]\,ds,\quad 
    \Delta \bU(\bX_*,\bX,\tau) := \bU\left(\bX_*+\bX,\tau\right)
    -\bU\big(\bX_*,\tau\big).
	\ee
The function $\bX_*(\tau)$ represents the relative separation between new and old scaling centers, which is prescribed by the initial value problem
	\be
	\label{eq3_8}
    \dfrac{d\bX_*}{d\tau} = \bU\left(\bX_*,\tau\right),\quad
    \bX_*(0) = \bX_0 := \frac{\widetilde{\bx}_0-\bx_0}{\ell}.
	\ee
Notice that $\bX_* = (\widetilde{\bx}_*-\bx_*)/\ell$ at the corresponding times. The transformations (\ref{eq3_6})--(\ref{eq3_8}) depend on the vector parameter $\bX_0 \in \mathbb{R}^3$, and map solutions $\bU(\bX,\tau)$ to new solutions $\widetilde{\bU}(\bX,\widetilde{\tau})$ of system (\ref{eq:newsystem_inviscid}). 
Thus, they represent the second nonlinear symmetry of the rescaled system. We call it the \textit{hidden translation symmetry}.\\

The five symmetries of the rescaled system, namely the three linear explicit ones and the two nonlinear hidden ones are summarized in Table~\ref{table:hiddensymmetries}. 
The table only applies to the inviscid system (\ref{eq:newsystem_inviscid}), which describes the proper dynamics in the inertial range,  with vanishing viscous and forcing terms. Upon suitable transformations of the rescaled forcing and dissipative terms (not shown), all five  symmetries can however be extended to the full system~(\ref{eq:newsystem}). 
\begin{table}[t]
 \caption{Symmetries of the dynamically rescaled Euler system (\ref{eq:newsystem_inviscid}) considered in $\mbR^3$. Rotation and parity symmetries require the  functional $\mathcal{A}[\cdot]$ to be rotation and parity invariant.}
\label{table:hiddensymmetries}
\begin{tabular}{lllll}
\hline\hline\\[-8pt]
			 		 &  parameters\quad			    & $\tau \mapsto \hs$  		& $\bX \mapsto \hs$ 	& $\bU \mapsto  \hs$   \\[3pt]
\hline
\\[-8pt]
Rotation &$\bO\in \mathrm{SO}(3) $ &$ \tau$  &$ \bO\bX$ &$ \bO\bU $  \\[3pt]
Parity & &$ \tau$  &$ -\bX$ &$ -\bU$    \\[3pt]
Time translation & $\Delta\tau \in \mbR$ &$ \tau+\Delta\tau$  &$ \bX$ &$  \bU$  \\[3pt]
Hidden translation & $\bX_0 \in \mathbb{R}^3$		& $\widetilde{\tau}$ in Eq. (\ref{eq3_7}) & $ \bX$	&$ 	\widetilde{\bU}$ in Eq. (\ref{eq3_6})\
\\[3pt]
Hidden scaling & $\lambda >0 $		& $\tau'$ in Eq. (\ref{eq3_5}) & $ \bX$	& 		$\bU'$ in Eq. (\ref{eq3_4})    
\\[3pt]
\hline\hline
\end{tabular}
\vspace*{-4pt}
\end{table}

\section{Relation between hidden and original symmetries}
\label{sec:four}
As seen from direct comparison between  Tables \ref{table:symmetries} \& \ref{table:hiddensymmetries}, the three linear symmetries are direct analogues to those of the original system. In other words, the  dynamical rescaling \emph{preserves}  rotation, parity and time-translation symmetries.
The status of the  nonlinear hidden symmetries is more intricate. On the one hand, the remaining symmetries of the original system is an eight-parameter group composed of spatio-temporal scaling (two), space-translation (three) and Galilean symmetries (three parameters).
On the other hand, the group of hidden symmetries is a four-parameter group, with three parameters for hidden translation, and one for hidden scaling.
We now show that the hidden  symmetries essentialize a "collapse" of  spatio-temporal scaling, space-translation and Galilean symmetries of the original system, with important implications for turbulence.

\paragraph{Fusing of the spatio-temporal scaling.} First, let us see how the hidden scaling symmetry (\ref{eq3_4})--(\ref{eq3_5}) is linked to the space-time scaling symmetries of the original system. The latter have the form (see Table \ref{table:symmetries})
	\begin{equation}
	\label{eq:spacetimeeuler}
	t,\;\mathbf{x},\;\mathbf{u} \ \ \mapsto \ \
	\alpha t, \;\lambda \bx, \; (\lambda/\alpha) \bu \quad
	\mathrm{for} \quad 
	\alpha, \lambda > 0.
	\end{equation}
One can check that, for any symmetry (\ref{eq:spacetimeeuler}), the corresponding  rescaled field $\bU(\bX,\tau)$ transforms according to the hidden scaling symmetry (\ref{eq3_4}) with the same parameter $\lambda$ and independent of $\alpha$; see Appendix \ref{app:C} for details. The fact that the temporal scaling with any $\alpha > 0$ does not affect the rescaled field $\bU(\bX,\tau)$ implies that our dynamical rescaling \emph{projects} the two-parameter family of space-time symmetries (\ref{eq:spacetimeeuler})  onto the one-parameter family of hidden scaling symmetries. 

Relations (\ref{eq:spacetimeeuler}) can be written in the form (\ref{eqI_1}) for $\alpha = \lambda^{1-h}$.   In particular, the statistical self-similarity of $\mathbf{u}(\mathbf{x},t)$ with any $h$ (for example, $h=1/3$ in Kolmogorov 1941 theory) translates into the hidden scaling symmetry for the rescaled solution $\bU(\bX,\tau)$. Furthermore, since the hidden symmetry does not depend on $h$, it is restored in solutions combining  scale-invariant parts with different exponents $h$.  This property makes the hidden scaling symmetry compatible with the multifractal framework of entangled scaling symmetries mentioned in Introduction: intermittent solutions restore the hidden scale-invariance in the statistical sense for the rescaled fields $\bU(\bX,\tau)$, while all scaling symmetries (\ref{eqI_1}) are broken in original variables. 
One can see this property as a tradeoff: by loosing the information on exponents $h$ we restore the self-similarity.
An explicit solvable shell-model example with this effect was recently presented in \cite{mailybaev2021solvable}.

\paragraph{Fusing of  space-translation and Galilean symmetries.}
{\color{black} Similarly, let us consider the six-parameter  family combining spatial translations with Galilean symmetry (see Table~\ref{table:symmetries}) as
	\begin{equation}
	\label{eq4_2}
	t,\;\mathbf{x},\;\mathbf{u} \ \ \mapsto \ \
	 t, \;\mathbf{x}+\Delta \mathbf{x}+t\mathbf{u}_0,\;\mathbf{u}+\mathbf{u}_0 
	\quad \mathrm{for} \quad \Delta \mathbf{x},\; \mathbf{u}_0 \in \mathbb{R}^3.
	\end{equation}
This transformation changes the Lagrangian trajectories and their initial points as
\be
    \label{eq4A_1}
    \bx_*(t) \mapsto 
    \bx_*(t)+\Delta \bx+t\bu_0,\quad
    \bx_0 \mapsto 
    \bx_0+\Delta \bx.
\ee
The straightforward consequence is that the field $\Delta\bu_\ell(\bx_*(t),\bX,t)$ defined in (\ref{eq_resc_sp}) and computed along the Lagrangian trajectory $\bx_*(t)$ is invariant with respect to the changes (\ref{eq4_2}) and (\ref{eq4A_1}). Therefore, the rescaled field $\bU(\bX,\tau)$ in Eq. (\ref{eq:newvariables}) remains intact. 

In addition to (\ref{eq4A_1}), let us consider a different Lagrangian trajectory $\widetilde{\bx}_*(t)$ with the initial point $\widetilde{\bx}_0 = \bx_0-\Delta\bx$ and the corresponding rescaled field $\widetilde{\bU}(\bX,\widetilde{\tau})$ given by Eqs.~(\ref{eq3_6})--(\ref{eq3_8}) of the hidden translation symmetry with $\bX_0 = -\Delta\bx/\ell$. Then, the symmetry transformation (\ref{eq4_2}) yields
\be
    \label{eq4A_1b}
    \widetilde{\bx}_*(t) \mapsto 
    \widetilde{\bx}_*(t)+\Delta \bx+t\bu_0,\quad
    \widetilde{\bx}_0 \mapsto 
    \widetilde{\bx}_0+\Delta \bx = \bx_0.
\ee
We see that $\widetilde{\bU}(\bX,\widetilde{\tau})$ is the rescaled field obtained after the symmetry transformation (\ref{eq4_2}) if we insist on keeping the same initial point $\bx_0$ for the Lagrangian trajectory. Through this relation, the hidden translation symmetry becomes the rescaled form of symmetries (\ref{eq4_2}) under the extra condition that the initial point of the Lagrangian scaling center remains fixed. It follows, in particular, that Galilean transforms do not affect the rescaled field $\bU(\bX,\tau)$, implying that the six-parameter family (\ref{eq4_2}) is projected by our dynamical rescaling onto the three-parameter family of hidden translation symmetries.}

In the context of turbulence, the role of the hidden translation symmetry is to cope with the sweeping effect by introducing the so-called quasi-Lagrangian formulation (see, e.g. \cite{biferale1999multi}): the reference frame moving with a Lagrangian particle removes undesirable effects of large-scale motion on small-scale statistics.\\

The connection between original and hidden symmetries is the crucial outcome of our dynamical rescaling of the original NS equations. We showed that the rescaling {\it preserves} some of the original symmetries (rotation, parity and temporal translation), while the remaining ones (spatio-temporal scaling, Galilean and space-translation) are {\it projected} onto  the four-parameter family of hidden symmetries.
Namely, applying an arbitrary combination of Galilean transform and temporal scaling to a flow $\bu(\bx,t)$ yields exactly the same field $\bU(\bX,\tau)$. As a consequence, new (weaker) hidden symmetries emerge, which are compatible with the intermittent turbulent statistics. Notice that this point of view can be formulated rigorously as a quotient construction within a more general group-theoretical framework of dynamical systems~\cite{mailybaev2020hiddena}. 

\section{Kolmogorov multipliers}
\label{sec:five}
As an application, we now expose a relation between the hidden scaling symmetry and the intrinsic refined self-similarity hypothesis -- the scale-invariance of Kolmogorov multipliers. Let $\bu(\bx,t)$ be the velocity field for the NS system. Consider a Lagrangian trajectory $\bx_*(t)$ defined by an initial point $\bx_0 = \bx_*(0)$, and the corresponding rescaled field  $\bU(\bX,\tau)$ for a given $\ell$. In terms of this field, we introduce the rescaled multipliers as
\be
    \label{eq5_N1}
    W_{ij,k}(\bx_0,\tau;\ell,\lambda_1, \lambda_2)
	:= \dfrac{U_k(\lambda_1\mathbf{e}_i,\tau)}{U_k(\lambda_2\mathbf{e}_j,\tau)},
\ee
where we explicitly indicated the dependence upon the trajectory via its initial point $\bx_0$ and on the scale $\ell$.
Using expressions (\ref{eq:turbulentmultipliers}), (\ref{eq_resc_sp}), (\ref{eq:newvariables}) and (\ref{eq5_N1}), one obtains
\be
    \label{eq4_3}
	W_{ij,k}(\bx_0,\tau;\ell,\lambda_1, \lambda_2)
	= w_{ij,k}\big(
	\bx,t;\ell_1, \ell_2
	\big),
\ee
where 
\be
    \label{eq5M_1}
    \bx = \bx_*(t),\quad 
    \ell_1 = \lambda_1\ell, \quad
    \ell_2 = \lambda_2\ell,
\ee
and $\tau$ is related with $t$ by the integral in (\ref{eq:newvariables}). 

Assume now that the velocity field $\bu(\bx,t)$ corresponds to a state of fully developed homogeneous turbulence, and the scales $\ell$, $\ell_1$ and $\ell_2$ belong to the inertial interval. Let us denote by $\langle \cdot \rangle_\tau$ an average with respect to time $\tau$, which we estimate for an arbitrary observable, prescribed by  a real function $f(w)$ with $w = W_{ij,k}(\bx_0,\tau;\ell,\lambda_1, \lambda_2)$. By homogeneity, we extend this average to all initial points $\bx_0$, denoted as $\langle \cdot \rangle_{\bx_0,\tau}$:
\be
    \label{eq5N_2}
    \left\langle f\left(W_{ij,k}\right)
	\right\rangle_\tau
    = \left\langle f\left(W_{ij,k}\right)
    \right\rangle_{\bx_0,\tau}.
\ee
The relation $\bx = \bx_*(t)$ and the second expression in (\ref{eq:newvariables}) define a one-to-one correspondence between the points $(\bx,t)$ and $(\bx_0,\tau)$. We compute the Jacobian of the map $(\bx,t) \mapsto (\bx_0,\tau)$ as
\be
    \label{eq5N_3}
   \det \left[\frac{\partial (\bx_0,\tau)}{\partial(\bx,t)}\right]
    = 
    \det \left[\frac{\partial (\bx_0,\tau)}{\partial(\bx_0,t)}
    \right]
    \det \left[
    \frac{\partial (\bx_0,t)}{\partial(\bx,t)}
    \right]
    = \left.\frac{\partial\tau}{\partial t}\right|_{\bx_0}
    = \frac{1}{T_\ell(\bx,t)},
\ee
where the third determinant is unity because of incompressibility. Hence, the right-hand side in (\ref{eq5N_2}) transforms into an average $\langle \cdot \rangle_{\bx,t}$ with respect to $\bx$ and $t$ as
\be
    \label{eq5N_4}
   \left\langle f\left(W_{ij,k}\right)
	\right\rangle_{\bx_0,\tau}
    = \left\langle f\left(w_{ij,k}\right) J_\ell(\bx,t)
	\right\rangle_{\bx,t},
	\quad
	\text{with}\quad
	J_\ell(\bx,t)=\dfrac{1/T_\ell(\bx,t)}{\av{1/T_\ell(\bx,t)}_{\bx,t}},
\ee
where we substituted $W_{ij,k}$ by $w_{ij,k}$ from (\ref{eq4_3}), and the pre-factor $1/T_\ell$ from (\ref{eq5N_3}) accounts for the change of coordinates.
We now use the natural hypothesis that the quantities $T_\ell(\bx,t)$ and $w_{ij,k}\big(\bx,t;\ell_1, \ell_2\big)$ become statistically independent when $\ell \gg \ell_1,\ell_2$, {\it i.e.}, when the time scale is evaluated at a much larger scale than the multiplier. In this case, we can factorize the average 
\be
    \label{eq5N_5}
    \left\langle  f\left(w_{ij,k}\right)J_\ell(\bx,t)\right\rangle_{\bx,t} 
    \approx 
    \left\langle  f\left(w_{ij,k}\right)\right\rangle_{\bx,t} 
    \left\langle J_\ell(\bx,t) \right\rangle_{\bx,t}
    \quad
    \textrm{for}\quad
    \ell \gg \ell_1,\ell_2,
\ee
where $\left\langle J_\ell(\bx,t) \right\rangle_{\bx,t} = 1$ by its definition in (\ref{eq5N_4}).
Combining (\ref{eq5N_4}) with (\ref{eq5N_5}) and writing it in reverse order with explicit arguments, we have
\be
    \label{eq5N_6}
    \left\langle f\left(w_{ij,k}
    \left(
	\bx,t;\ell_1, \ell_2
	\right)
	\right)
	\right\rangle_{\bx,t}
	\approx
   \left\langle f\left(W_{ij,k}
   (\bx_0,\tau;\ell,\lambda_1, \lambda_2)\right)
	\right\rangle_{\bx_0,\tau}
	\quad
    \textrm{for}\quad
    \ell \gg \ell_1,\ell_2.
\ee
This formula expresses the average of any single-time observable depending on the Kolmogorov multiplier $w_{ij,k}$ as the average in rescaled variables. 

{\color{black}We now assume that the hidden scaling symmetry is restored in the inertial interval, namely,
\be
    \label{eq_HSI}
    \bU(\cdot,\tau) 
    \ \overset{\textrm{law}}{=}\
    \bU'(\cdot,\tau') 
\ee
in a statistical sense, where the field $\bU'(\bX,\tau')$ given by Eqs.~(\ref{eq3_4})--(\ref{eq3_5}) corresponds to a different zooming scale $\ell'$ from the inertial interval. 
In other words, hidden scaling symmetry means that the statistics of $\bU$ do not depend on the zooming scale $\ell$.
In this case any statistical property of the rescaled field, like the right-hand side of (\ref{eq5N_6}) expressing the average for an arbitrary function of a multiplier, is independent of the choice of $\ell$. 
Recalling that $\ell_1 = \lambda_1\ell$ and $\ell_2 = \lambda_2\ell$, we conclude that the expression in (\ref{eq5N_6}) depends on the scales $\ell_1$ and $\ell_2$ only through their ratio $\ell_2/\ell_1$.} This is exactly the scale invariance property of the Kolmogorov multipliers, which we obtained as a consequence of the hidden scaling symmetry. Notice the crucial role of incompressibility in our derivation, which yields a simple expression for the Jacobian (\ref{eq5N_3}); it does not extend to compressible flows. 

\section{Hidden scale-invariance: numerical experiments}
\label{sec:six}
We now  present some statistical analysis of large direct numerical simulations, in order to test the hidden scaling symmetry for fully developed homogeneous turbulence. The hypothesis of hidden scale invariance formulated in (\ref{eq_HSI}) implies that the statistics of rescaled fields $\bU(\bX,\tau)$ are not dependent upon the scale $\ell$ in the inertial interval. As  shown in \S\ref{sec:three}, this symmetry is equivalent to the invariance of statistical properties with respect to transformations (\ref{eq3_4})--(\ref{eq3_5}) depending on the scaling ratio $\lambda$.

\paragraph{Statistical methodology.}
We use  highly resolved data available from a massive  pseudo-spectral numerical simulation of the NS system using $4096^3$ collocation points in a triply periodic domain. Properties of this data are fully documented in \cite{homann2007impact,bitane2013geometry} and relevant physical parameters are summarized in Table \ref{table:dns}.  In the  steady state, the inertial range spans more than two decades of spatial scales with $L/\eta \sim 440$. 
 To compute statistical properties of the flow, we use Monte-Carlo sampling relying on $5 \times 10^7$ random points uniformly distributed on  the computing  domain $[0,2\pi]^3$ and four different turbulent snapshots separated by the times of order $T_L/2$.

\begin{table}
\caption{Physical parameters for the $512^3$ simulations displayed in Figs. \ref{fig:non-Gaussian} \& \ref{fig:multipliers}, and for the $4096^3$ simulation used in \S\ref{sec:six}. We define the integral-scale velocity $U=\av{\bu^2/3}^{1/2}$,  dissipation  rate $\epsilon=\av{\nu ||\nabla \bu||^2}$, spatial scale  $L=U^3/\epsilon$, temporal scale $T_L=L/U$, Reynolds number $Re=UL/\nu$ and dissipative scale $\eta=2\pi (\nu^3/\epsilon)^{1/4}$. The coefficient $c_2>0$ indicates presence of intermittency. It is obtained by fitting the structure function exponents $\zeta_p$ for $1 \le p \le 8$ with quadratic refined-self-similarity formula $\zeta_p = c_1p - c_2p^2/2$, $c_1 = 3 c_2/2+1/3$, as in  Fig.~\ref{fig:non-Gaussian}.}
\label{table:dns}
\begin{tabular}{lrrrrrrrr}
\hline\hline
\\[-8pt]
Run	 & $ L$ &$U$  & $T_L$   & $\epsilon$  & $\eta$  &  Re &$L/\eta$& $c_2$\\[3pt]
\hline
\\[-8pt]
$512^3$	 &1.5  &0.85  & 1.8   & 0.41  & 0.026  &  2600 &58& $0.026 \pm 0.003$\\[3pt]
$4096^3$&  2.0  & 0.19  & 10.4   & 0.0036  & 0.0046  &  39000 &440& $0.033 \pm 0.003$\\[3pt]
\hline
\hline
\end{tabular}
\vspace*{-4pt}
\end{table}

It is convenient to express the statistical properties of the rescaled fields, whose definitions involve Lagrangian scaling centers, in terms of Eulerian averages by following steps similar to \S\ref{sec:five}.
Consider an arbitrary single-time observable $\Phi[\bU(\cdot,\tau)]$ defined for the rescaled field $\bU(\bX,\tau)$. We introduce the corresponding observable in terms of the original field as
\begin{equation}
	\label{eq6_2}
\varphi_{\bx,\ell}[\bu(\cdot,t)] :=
\Phi\left[
\frac{\Delta\bu_\ell(\bx,\cdot,t)}{A_\ell(\bx,t)}\right],
\end{equation}
depending on position $\bx$ and scale $\ell$.
Using the expressions (\ref{eq_resc_sp}) and (\ref{eq:newvariables}), one can see that
\begin{equation}
	\label{eq6_3}
\Phi[\bU(\cdot,\tau)] = \varphi_{\bx,\ell}[\bu(\cdot,t)]
\quad \textrm{for}\quad
\bx = \bx_*(t),\quad 
\tau = \int_0^{t}\frac{ds}{T_\ell(\bx_*(t),t)}.
\end{equation}
Thus, $\varphi_{\bx,\ell}$ represents the observable $\Phi$ expressed in terms of the original variables. Similarly to the derivation of Expression (\ref{eq5N_4}) for the homogeneous statistics of incompressible flow, we have
\begin{equation}
	\label{eq6_4}
   \left\langle \Phi[\bU(\cdot,\tau)]
	\right\rangle_\tau
    =
    \left\langle \varphi_{\bx,\ell}[\bu(\cdot,t)\;]J_\ell(\bx,t)
	\right\rangle_{\bx,t},
    \quad
    J_\ell(\bx,t)=
    \frac{1/T_\ell(\bx,t)}{
	\left\langle 1/T_\ell(\bx,t)
	\right\rangle_{\bx,t}}.
\end{equation}
This formula provides a practical way to compute averages involving observables of the rescaled field, but without performing the actual transformation to new variables.

Specifically, we use $\Phi$ as an averaged indicator function depending on two real parameters $U_\parallel$ and $\lambda$, and defined for an arbitrary field $\bV(\bX)$ as
\begin{equation}
	\label{eq6_5}
	 \Phi[\bV] =	
			\av{\delta\left(U_\parallel-V_\parallel(\lambda\bX)\right)}_S
		\quad \text{with} \quad  V_\parallel(\bX) = \bV(\bX) \cdot \bX/\|\bX\|.
\end{equation}
Here $\delta$ is the Dirac delta-function, and $\langle\cdot\rangle_{S}$ denotes the averaging over the  unit spherical shell $\|\bX\| = 1$.
Then, using (\ref{eq6_4}), we express the average 
\be
	\label{eq6_6}
    \rho_{\ell,\lambda}(U_\parallel) := \left\langle \Phi
    [\bU(\cdot,\tau)]	\right\rangle_\tau=\left\langle \delta\left(U_\parallel-\left(\dfrac{\Delta \bu_\ell(\bx,\lambda\bX,t)}{A_\ell(\bx,t)}\right)_\parallel\right)
    J_\ell(\bx,t)
    \right\rangle_{\bx,t,S},
\ee
which represents the ``hidden'' distribution (PDF) of the parallel rescaled field $U_\parallel(\bX,\tau) = \bU(\bX,\tau) \cdot \bX/\|\bX\|$, constructed with a given zooming scale $\ell$ and observed at a rescaled distance $\|\bX\| = \lambda$. Statistical hidden scale invariance implies that the distribution $\rho_{\ell,\lambda}(U_\parallel)$ does not depend on $\ell$, provided that the scales $\ell$ and $\lambda \ell$ belong to the inertial interval. 

\paragraph{Prescribing the functional $\mA$.}
Our theory leaves a freedom for choosing the functional $\mA[\cdot]$ for our definitions in (\ref{eq2_4}). However, we do not expect that every choice  will cope equally well with finite-size effects: some operators might be better than others for revealing statistical signatures of hidden scale invariance within a rather large but still finite inertial range.
We tested three different prescriptions: 
\be
\label{eq6_A}
(i)\ \ \mA[\bV]= \sqrt{\av{\|\bV\|^2}_S},\quad
(ii)\ \ \mA[\bV]= \left|V_\parallel(\mathbf{e})\right|,\quad
(iii)\ \ \mA[\bv]=\left|\sqrt[3]{\av{||\bV||^2  V_\parallel}_S}\right|,
\ee
where $V_\parallel(\bX) = \bV(\bX) \cdot \bX/\|\bX\|$, and the average is over the unit spherical shell $S$.
The first choice is based on local (unit sphere) energy estimate,  which we prefer to the unit ball for numerical reasons. The second choice is motivated by the Kolmogorov multipliers written in the form (\ref{eq5_N1}), with a fixed (picked randomly) unit vector $\mathbf{e}$.  The motivation for the third choice comes from the Duchon-Robert (DR) local dissipation law \cite{duchon2000inertial} relevant for developed turbulence \cite{kuzzay2015global,eyink2002local}, in which the quantity $\|\delta \bu\|^2 \delta u_\parallel$ is related to the local energy dissipation.

\begin{figure}[ht]
\centering
\hspace{5mm} (a) \hspace{61mm} (b)
\\
\includegraphics[width=0.49\textwidth]{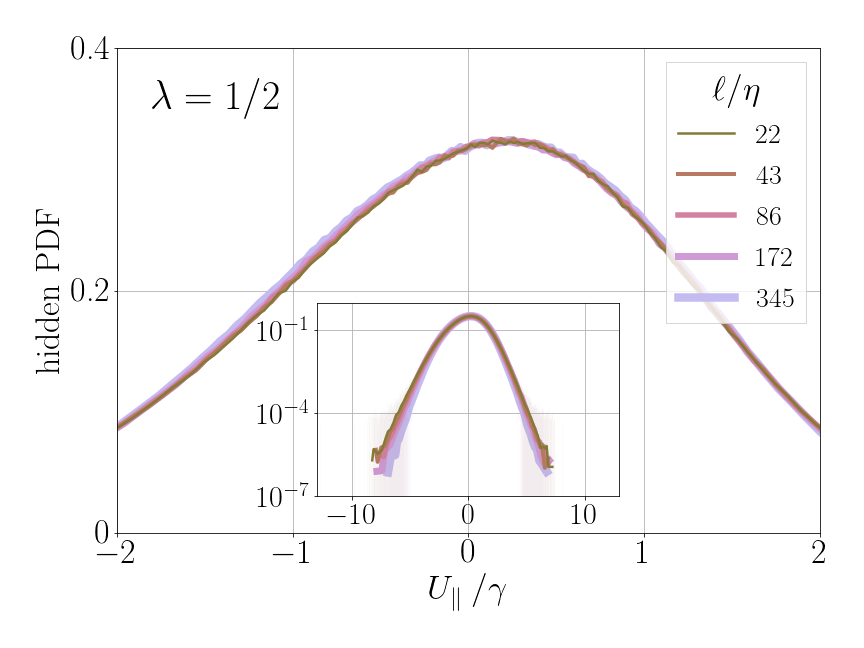}
\includegraphics[width=0.49\textwidth]{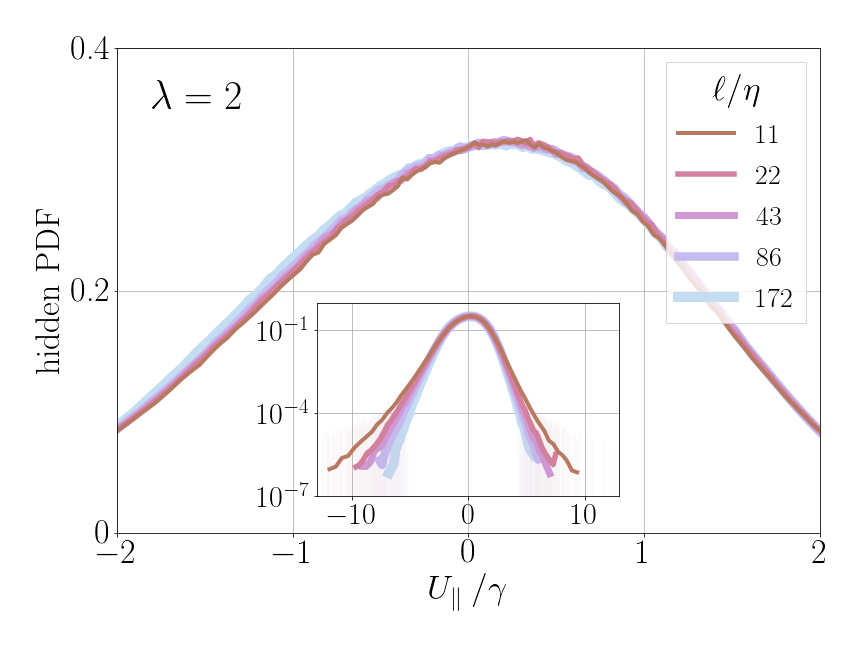}
\\
\hspace{5mm} (c) \hspace{61mm} (d)
\\
\includegraphics[width=0.49\textwidth]{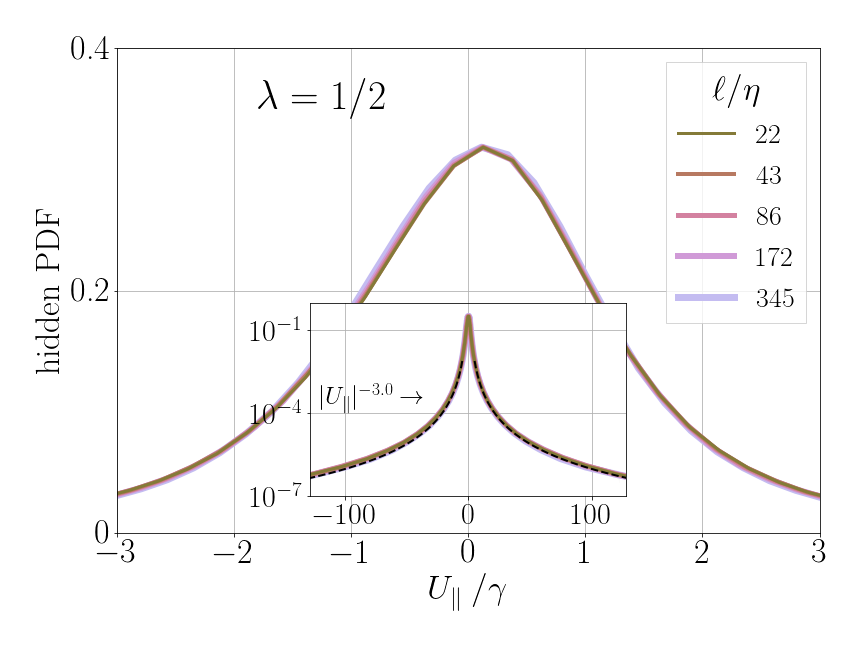}
\includegraphics[width=0.49\textwidth]{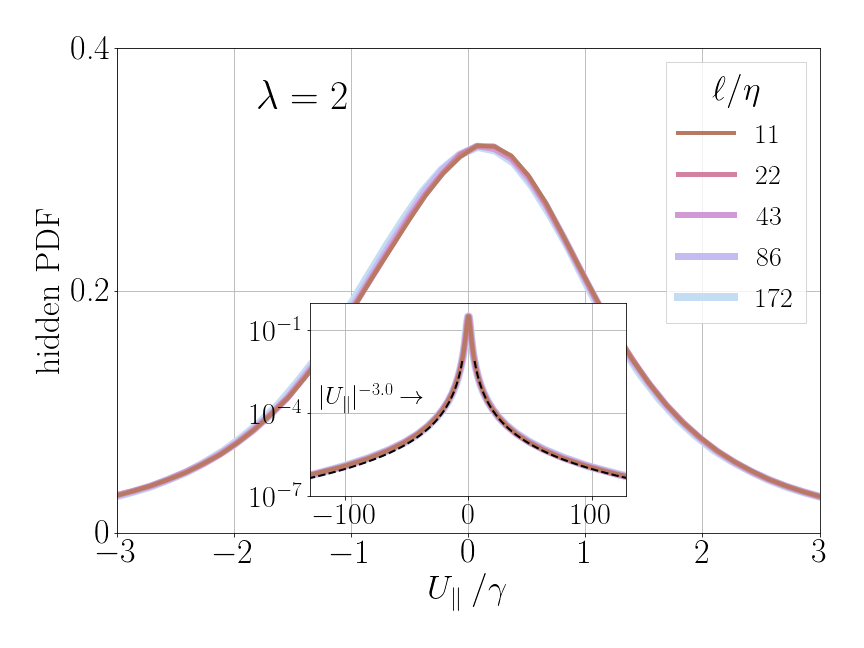}
\\
\hspace{5mm} (e) \hspace{61mm} (f)
\\
\includegraphics[width=0.49\textwidth]{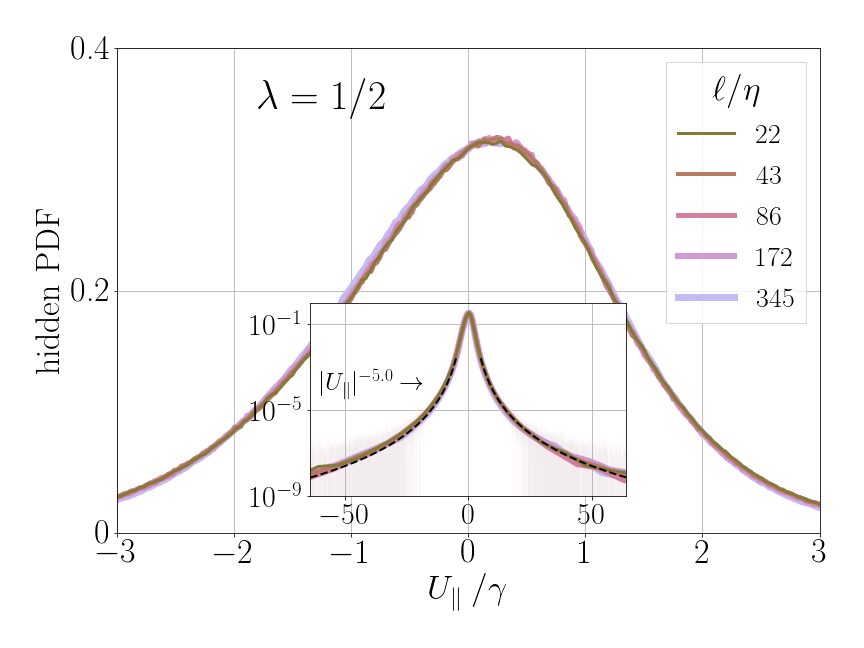}
\includegraphics[width=0.49\textwidth]{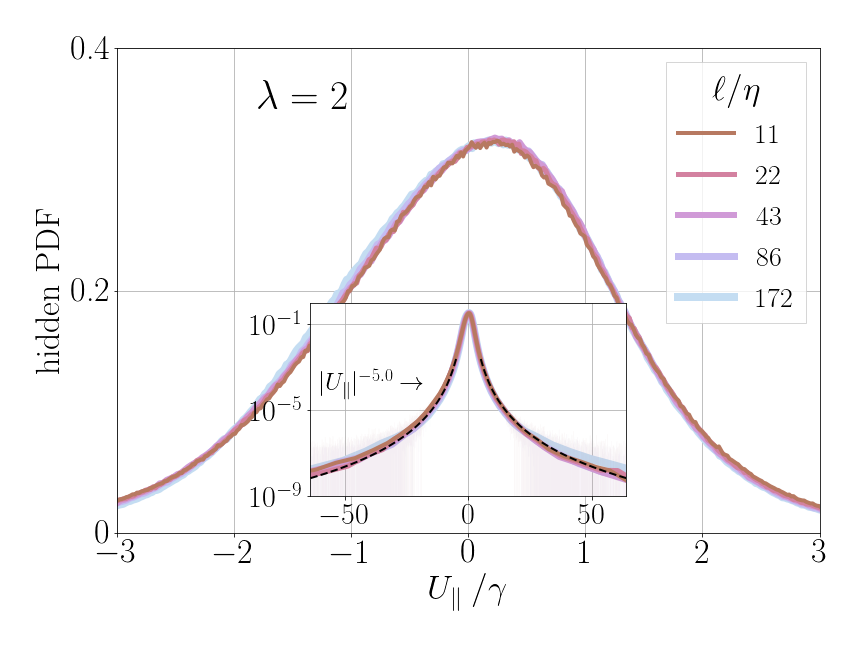}
\caption{Normalized hidden PDF $\rho_{\ell,\lambda}(U_\parallel)$ of the parallel rescaled field $U_\parallel$ for different zooming scales $\ell$; insets show tails of the same graphs in vertical logarithmic scale. The panels (a,b), (c,d) and (e,f) correspond, respectively, to the choices ($i$), ($ii$) and ($iii$) from (\ref{eq6_A}) for the operator $\mA$. The (dashed) algebraic fits in the insets are empirical, indicative of the tail behaviors.
Similarly to Fig.~\ref{fig:multipliers}, we use the scale factor $\gamma=1/{(\pi \max \rho_{\ell,\lambda})}$ to normalize the PDF as $U_\parallel,\rho_{\ell,\lambda} \mapsto U_\parallel/\gamma,\gamma \rho_{\ell,\lambda} $.  }
\label{fig:HiddenDV2}
\end{figure}

\paragraph{Results.}
Figure \ref{fig:HiddenDV2} summarizes the numerical results for the three respective choices of the amplitude functional (\ref{eq6_A}).
These figures show the hidden PDF $\rho_{\ell,\lambda}(U_\parallel)$ for different values of the zooming scale $\ell$ and fixed $\lambda = 1/2$ (left) or $\lambda = 2$ (right). 
All panels of Fig. \ref{fig:HiddenDV2} demonstrate a remarkable collapse of PDFs in their central part for over a decade of scales $\ell$ from the inertial interval. 
These results substantiate our hypothesis that the hidden scale invariance is restored statistically in the fully developed turbulence. 

We now draw attention to the insets of Fig. \ref{fig:HiddenDV2}. For the choices ($ii$) and  ($iii$) of the amplitude functional (\ref{eq6_A}), the insets in panels (c--f) show an accurate collapse of PDF tails. In the case ($ii$), the tails are dominated by small denominators $A_\ell$ in (\ref{eq6_2}), leading to Cauchy distributions (shown by dashed lines) in the context of multipliers~\cite{chen2003kolmogorov}. It is likely that small denominators $A_\ell$ are also responsible for the shape of tails in the case ($iii$); we showed by dashed lines an empirical power law $\propto |U_\parallel|^{-3}$.
On the contrary, the choice ($i$) yields a considerable separation of the tails in the insets of panels (a,b). A possible interpretation is that small denominators $A_\ell$ are inhibited by the choice of the local energy estimate in (\ref{eq6_A}). In this case the tails are dominated by large velocity differences in the numerator of Eq. (\ref{eq6_2}), and this may retard the convergence of PDFs.

We also mention an empirical observation that statistical averages (\ref{eq6_A}) with the removed Jacobian factor $J_\ell$  still lead to the apparent scale invariance, but with different self-similar profiles. For both  choices {\it (ii)} and {\it (iii)}, this  reflects in shallower tails; in the former case, one recovers the power-law decay $\propto |U_\parallel|^{-2}$ compatible with the Cauchy distributions of Fig.~\ref{fig:multipliers}.

We conclude with a methodological remark that all PDFs in Figs. \ref{fig:non-Gaussian}, \ref{fig:multipliers} and \ref{fig:HiddenDV2} are normalized. In particular  Fig. \ref{fig:HiddenDV2} represents the PDF obtained by introducing the scaling factors $\gamma=1/({\pi \max \rho_{\ell,\lambda}})$, hereby transforming 
$U_\parallel \mapsto U_\parallel/\gamma$ and $\rho_{\ell,\lambda} \mapsto \gamma \rho_{\ell,\lambda}$. 
A similar transform 
was used in Fig.~\ref{fig:multipliers}.
Without this rescaling the collapses deteriorate considerably in the cases {\it (ii)} and {\it (iii)}, as well as in Fig.~\ref{fig:multipliers}. We connect this feature to the emergence of heavy-tailed statistics, which produce data-size-dependent (irrelevant) normalizing factors when computing the empirical PDFs. 
In the case {\it (i)} though, quality of the collapse is not affected for moderate $\lambda \lesssim 1$. 
We defer to future work a detailed study of these non-normalized statistics.

\section{Concluding remarks}
\label{sec:seven}

We have exposed a general dynamical rescaling of the NS system,  featuring a change to a quasi-Lagrangian reference frame with a subsequent spatial scaling and suitably defined clock associated to the local flow intensity. From the point of view of symmetries, the transformation preserves the usual rotation, parity and time-translation symmetries of the NS system, but projects the remaining ones onto a reduced parametric set of \emph{hidden symmetries} for the equations of motion governing the rescaled fields. This way, the classical two-parameter family of spatiotemporal scalings, $t,\bx,\bu \to  \lambda^{1-h}t, \lambda  \bx, \lambda^h \bu$, fuses into a one-parameter hidden scaling symmetry independent of the exponent $h$. 
Similarly, translations and Galilean transforms are projected onto the three-parameter hidden translation symmetry. 

We conjecture that hidden symmetries reflect statistical properties of velocity fields in the vanishing viscosity limit. 
For the fully developed turbulence, this means that hidden scaling symmetry may be restored statistically in the inertial interval, although the scaling symmetries in the usual sense are statistically broken by intermittency. 
This possibility substantiates the intrinsic version of the refined self-similarity formulated by Kolmogorov in his 1962 paper: The scale invariance of Kolmogorov multipliers follows from the hidden scale invariance.
Extending  to  multi-point statistics, hidden scale invariance implies statistical universality of the correlations between the multipliers, hereby substantiating the statistical modeling of  multifractality in terms of   multiplicative processes \cite{kahane1985chaos,friedrich2018multiscale,chevillard2019skewed}. There is no direct relation between the hidden symmetry and energy conservation (or anomalous dissipation) \cite{mailybaev2021solvable}, but the latter might impose restrictions on statistical properties for the universal quantities.



Performing statistical analysis of large direct numerical simulations, we provided the evidence that the hidden scaling symmetry is indeed restored in the inertial range of fully developed turbulence. We focused on the statistics of single-time observables, showing the scale invariance of longitudinal components of rescaled fields.
The hidden symmetries are also relevant and may be verified numerically in multi-time statistics, which remains a challenge for future work.

While we considered here the NS system, relevance of hidden symmetry to other types of intermittency is clearly a subject worth investigating: in the Burgers equation \cite{weinan1999asymptotic,bec2007burgers}, scalar transport \cite{gawedzki1995anomalous} or financial markets \cite{bouchaud2000apparent}, to name a few. We recall that our derivations in Section~\ref{sec:five} rely on the incompressibility property; As mentioned in \cite{mailybaev2020hiddena}, this may pose some limitations when extending our techniques to compressible models.

\appendix
\section{Appendix}
\subsection{Derivation of the dynamically rescaled NS equations}
\label{app:derivation}
Here we derive system (\ref{eq:newsystem}) from the NS equations under the dynamical rescaling by Eqs.~(\ref{eq_resc_sp}) and (\ref{eq:newvariables}).
For the sake of clarity, we use the notations $\nabla_{\bX}$ and $\Delta_{\bX}$ for the gradient and Laplace operators in new coordinates $\bX$. We focus on the first equation in (\ref{eq:newsystem}), since the incompressibility condition follows trivially. Using relations (\ref{eq2_4}) and (\ref{eq:newvariables}) we derive 
\be
\label{eqA_1}
    \partial_\tau \bU 
    = T_\ell\dfrac{d}{dt}\frac{\Delta\bu_\ell}{A_\ell}
    = \frac{T_\ell}{A_\ell}\dfrac{d
    \Delta\bu_\ell}{d t}
    -\frac{T_\ell\,\Delta\bu_\ell}{A_\ell^2}\dfrac{d A_\ell}{dt}
    = \widetilde{\bV}
    -\bU
    \left.\dfrac{\delta \mathcal{A}}{\delta\bV}\right|_{\Delta\bu_\ell(\bx_*(t),\cdot,t)}
    \widetilde{\bV},
\ee
where 
\be
\label{eqA_1a}
    \widetilde{\bV} = \frac{T_\ell}{A_\ell}\dfrac{d\Delta\bu_\ell}{dt},
\ee
with the arguments (omitted for simplicity) being $\Delta\bu_\ell(\bx_*(t),\bX,t)$, $A_\ell(\bx_*(t),t)$ and $T_\ell(\bx_*(t),t)$, and the derivative $d/dt$ computed along the Lagrangian trajectory at fixed $\bX$.
From the degree-1 homogeneity property (\ref{eq_A_hom}), we have $\mathcal{A}(\Delta\bu_\ell+\delta\bV) = A_\ell\mathcal{A}(\bU+\delta\bV/A_\ell)$ for $\bU = \Delta\bu_\ell/A_\ell$. This implies that the functional derivative $\delta\mathcal{A}/\delta\bV$ in (\ref{eqA_1}) can be evaluated at $\bU(\cdot,\tau)$ with the same result. This reduces equation (\ref{eqA_1}) to the form
\be
\label{eqA_1b}
    \partial_\tau \bU 
    = \widetilde{\bV}-\bU
    \left.\dfrac{\delta \mathcal{A}}{\delta\bV}\right|_{\bU(\cdot,\tau)}
    \widetilde{\bV}.
\ee

Using (\ref{eq_resc_sp}), (\ref{eq:Xstar}) and the shorthand notation $[g]_a^b=g(b)-g(a)$, we compute the derivative
\be
\label{eqA_2}
	  \dfrac{d}{dt} \Delta\mathbf{u}_\ell(\bx_*(t),\bX,t)
	  = \dfrac{d}{dt}\big[\bu(\bx,t)\big]_{\bx_*(t),t}^{\bx_*(t)+\ell\bX,t}
	  = \left[\dfrac{\partial\bu}{\partial t}\right]_{\bx_*,t}^{\bx_*+\ell\bX,t}
	  + \bu(\bx_*,t) \cdot \big[\nabla\bu\big]_{\bx_*,t}^{\bx_*+\ell\bX,t}.
\ee
Using the NS equation for the derivative $\partial\bu/\partial t$ and manipulating the convection terms, we have
\be
\label{eqA_3}
	  \dfrac{d}{dt} \Delta\mathbf{u}_\ell(\bx_*(t),\bX,t)
        = 
        -\big[\bu\big]_{\bx_*,t}^{\bx_*+\ell\bX,t} \cdot \nabla\bu(\bx_*+\ell\bX,t)
        +\left[
        -\nabla p
        +\mRe^{-1}{\Delta\bu } +\mathbf{f}
	\right]_{\bx_*,t}^{\bx_*+\ell\bX,t}.
\ee
Observing that the factor $T_\ell(\bx_*(t),t)/A_\ell(\bx_*(t),t)$ does not depend on spatial coordinates, and using Eqs.~(\ref{eq_resc_sp}), (\ref{eq2_4}), (\ref{eq:newvariables}) and (\ref{eqA_3}), we can write the right-hand side of expression (\ref{eqA_1a}) in terms of the new field $\bU(\bX,\tau)$. Taking into account the property $\bU(\mathbf{0},\tau) = \mathbf{0}$, and using the new pressure $P = p/A_\ell^2$, rescaled forcing term and Reynolds number defined in
(\ref{eq:newsystem_terms}), yields
\be
\label{eqA_4}
	  \widetilde{\bV}(\bX,\tau) 
        = \left[-\bU\cdot \nabla_\bX\bU
        -\nabla_\bX P
        +\mRe_{*\ell}^{-1}{\Delta_\bX\bU } +\mathbf{F}_{*\ell}
	\right]_{\mathbf{0},\tau}^{\bX,\tau}.
\ee
Combining this relation with (\ref{eqA_1b}), we derive the rescaled NS system (\ref{eq:newsystem}) with the linear operator (\ref{eq:LambdaU}). The pressure $p(\bx,t)$ of the NS equation with solenoidal forcing is determined by the Poisson equation $\Delta p = - \mathrm{Tr} (\nabla \bu)^2$ as a consequence of the incompressibility condition; see e.g. \cite{frisch1995turbulence}. Since $A_\ell$ in relations (\ref{eq:newvariables}) and $P = p/A_\ell^2$ does not depend on the coordinates $\bX$, the rescaled pressure $P(\bX,\tau)$ is given by the analogous Poisson equation $\Delta_\bX P = - \mathrm{Tr} (\nabla_\bX \bU)^2$ in new variables. 

\subsection{Derivation of the hidden symmetries}
\label{sec:hiddenscaleinvariance}
We here explicitly derive the transforms (\ref{eq3_4})--(\ref{eq3_5}) and
(\ref{eq3_6})--(\ref{eq3_8}) respectively defining hidden scaling and hidden translation symmetries.
\paragraph{Hidden scaling symmetry.}
Let $\bU(\bX,\tau)$ and $\bU'(\bX,\tau')$ be the two field given by expressions (\ref{eq_resc_sp}) and (\ref{eq:newvariables}) for the two scales $\ell$ and $\ell'$, respectively. From relation (\ref{eq_resc_sp}) it follows that 
	\begin{equation}
	\label{eqB1}
    \Delta\bu_{\ell'}(\bx,\bX,t) = \Delta\bu_{\ell}\left(\bx,\frac{\bX}{\lambda},t\right),\quad
    \lambda = \frac{\ell}{\ell'}.
	\end{equation}
Combining this identity with the definitions (\ref{eq2_4}) and (\ref{eq:newvariables}), we obtain the relation between the  fields $\bU$ and $\bU'$ as
\be
    \label{eqB2}
	\bU'(\bX,\tau')
    = \frac{\Delta\bu_{\ell'}(\bx_*(t),\bX,t)}{A_{\ell'}(\bx_*(t),t)}
    = \bU_\lambda( \bX,\tau)\; \dfrac{A_{\ell}(\bx_*(t),t)}{A_{\ell'}(\bx_*(t),t)},
\ee
with the shorthand $\bU_\lambda(\bX,\tau)=\bU(\bX/\lambda,\tau)$ as in Eq.~(\ref{eq3_5}).
Applying the operator $\mA$ on both sides of (\ref{eqB2}), and using properties (\ref{eq_A_hom}) and (\ref{eq2_star}),
we identify 
\be
    \label{eqB3}
   \mA[\bU_\lambda(\cdot,\tau)]=\dfrac{A_{\ell'}(\bx_*(t),t)}{A_{\ell}(\bx_*(t),t)}.
\ee 
Equalities (\ref{eqB2}) and (\ref{eqB3}) yield the relation (\ref{eq3_4}).
The relation (\ref{eq3_5}) between the rescaled times $\tau$ and $\tau'$ defined in (\ref{eq:newvariables}) follows from 
\be
    \label{eqB_4}
    d\tau'=\dfrac{dt}{T_{\ell'}(\bx_*,t)} = \underbrace{\dfrac{dt}{T_\ell(\bx_*,t)}}_{d \tau} 
     \underbrace{\dfrac{\ell}{\ell'}}_{\lambda} \underbrace{\dfrac{A_{\ell'}(\bx_*,t)}{A_\ell(\bx_*,t)}}_{\mA[\bU_\lambda(\cdot,\tau)]},
\ee
relying on the defining identities $T_\ell=\ell/A_\ell$ and $T_{\ell'}=\ell'/A_{\ell'}$ from Eq.~(\ref{eq2_4}) and Eq.~(\ref{eqB3}).

\paragraph{Hidden translation symmetry.}
Deriving the hidden translation symmetry (\ref{eq3_6})--(\ref{eq3_8}) follows similar steps. 
We define $\bU(\bX,\tau)$ and $\tilde \bU(\bX,\tilde \tau)$ as two rescaled fields for two distinct scaling centers $x_*(t)$ and $\tilde x_*(t)$ starting, respectively, from $\bx_0$ and $\tilde \bx_0$.
The corresponding fields (\ref{eq_resc_sp}) relate as
\be
    \label{eq:scaledtildefield}
    \Delta \bu_\ell(\tilde \bx_*,\bX,t)=\Delta \bu_\ell( \bx_*,\bX+\bX_*,t)-\Delta \bu_\ell( \bx_*,\bX_*,t),\quad  \bX_* = \dfrac{\tilde \bx_*-\bx_*}{\ell}.
\ee
From (\ref{eq:scaledtildefield}) with definitions  (\ref{eq_resc_sp}) and (\ref{eq:newvariables}), we now compute
\be
    \label{eqB_6}
    \tilde \bU(\bX,\tilde \tau ) = \dfrac{\Delta \bu_\ell(\tilde \bx_*(t),\bX,t)}{A_\ell(\tilde \bx_*(t),t)} =  \Delta \bU(\bX_*(\tau),\bX,\tau)\; \dfrac{A_\ell( \bx_*(t),t)}{A_\ell(\tilde \bx_*(t),t)},
\ee
with $\Delta \bU(\bX_*,\bX,\tau)=\bU(\bX_*+\bX,\tau)-\bU(\bX_*,\tau)$.
Applying the operator $ \mA$ to both sides of (\ref{eqB_6}) and using properties (\ref{eq_A_hom}) and (\ref{eq2_star}) yields 
\be
    \label{eqB_7}
    \mA\left[\Delta \bU(\bX_*(\tau),\cdot,\tau)\right]=\dfrac{A_\ell( \tilde \bx_*(t),t)}{A_\ell(\bx_*(t),t)}.
\ee
Together with (\ref{eqB_6}), this yields  
Eq.~(\ref{eq3_6}). 
In turn, the transformation~(\ref{eq3_7}) for the  rescaled times  follows from (\ref{eq2_4}), (\ref{eq:newvariables}) and (\ref{eqB_7}) as
\be
    d\tilde \tau=\dfrac{dt}{T_{\ell}(\tilde \bx_*,t)} = \underbrace{\dfrac{dt}{T_\ell(\bx_*,t)}}_{d \tau} 
     \underbrace{\dfrac{A_{\ell}(\tilde \bx_*,t)}{A_\ell(\bx_*,t)}}_{\mA[\Delta \bU(\bX_*(\tau),\cdot,\tau)]}.
\ee
Finally, one obtains the dynamics (\ref{eq3_8}) for the relative separation  in rescaled time $\tau$ by combining relations (\ref{eq_resc_sp}), (\ref{eq2_4}), (\ref{eq:Xstar}), (\ref{eq:newvariables}) and  (\ref{eq:scaledtildefield}) as
\be
    \label{eqB_10}
    \begin{array}{rcl}
    \displaystyle
    \dfrac{d\bX_*}{d\tau} 
    & = & \displaystyle
    \dfrac{T_\ell(\bx_*,t)}{\ell} \dfrac{d}{dt} \left(\tilde \bx_*(t) -\bx_*(t)\right)
    = 
    \frac{\bu(\tilde\bx_*,t) -\bu(\bx_*,t)}{A_\ell(\bx_*,t)}
    \\[12pt] & = & \displaystyle
    \dfrac{\Delta \bu_\ell \left(\bx_*,(\tilde \bx_*-\bx_*)/\ell,t\right)}{A_\ell(\bx_*,t)} = \bU(\bX_*,\tau). 
    \end{array}
\ee

\subsection{Dynamical rescaling and original scaling symmetries}
\label{app:C}

Let us consider the field $\bu'(\bx,t')$ obtained by the space-time scaling (\ref{eq:spacetimeeuler}). We express this field as 
\be
\label{eqC_0}
\bu'(\bx,t') = \dfrac{\lambda}{\alpha} \bu \left(\frac{\bx}{\lambda},t\right),\quad
t = \frac{t'}{\alpha}.
\ee
Given a Lagrangian trajectory $\bx_*(t)$ of the  field $\bu(\bx,t)$, one obtains the corresponding Lagrangian trajectory $\bx'_*(t')$ of the  field $\bu'(\bx,t')$ from the relation 
\be
\label{eqC_0b}
\bx'_*(t')=\lambda \bx_*(t),\quad
t = t'/\alpha.
\ee

Let us compute the dynamically rescaled field $\bU'(\bX,\tau')$ for velocity (\ref{eqC_0}), trajectory (\ref{eqC_0b}) and the same scale $\ell$.
Using  definition (\ref{eq_resc_sp}) with (\ref{eqC_0}) and (\ref{eqC_0b}), we have
\be
	\Delta\bu'_{\ell}(\bx_*'(t'),\bX,t') = \dfrac{\lambda}{\alpha} \Delta\bu_\ell\left(\bx_*(t),\frac{\bX}{\lambda} ,t\right).
\ee
Combining this expression with definition (\ref{eq:newvariables}) for $\bU'(\bX,\tau')$ and $\bU(\bX,\tau)$, yields
\be
    \label{eqC_1}
    \begin{array}{rcl}
	\bU'(\bX,\tau') & = &
	\displaystyle
	\dfrac{\Delta\bu'_{\ell}(\bx_*'(t'),\bX,t')}{A'_\ell(\bx_*'(t'),t')}= \dfrac{\lambda}{\alpha}
	\dfrac{A_\ell(\bx_*(t),t)}{A'_\ell(\bx_*'(t'),t')}
	\dfrac{\Delta\bu_{\ell}(\bx_*(t),\bX/\lambda,t)}{A_\ell(\bx_*(t),t)}
	\\[15pt] & = & \displaystyle 
	\dfrac{\lambda}{\alpha}
	\dfrac{A_\ell(\bx_*(t),t)}{A'_\ell(\bx_*'(t'),t')}
	\bU\left(\frac{\bX}{\lambda},\tau\right)
	= \dfrac{\lambda}{\alpha}
	\dfrac{A_\ell(\bx_*(t),t)}{A'_\ell(\bx_*'(t'),t')}
	\bU_\lambda\left(\bX,\tau\right),
	\end{array}
\ee
with $\bU_\lambda(\bX,\tau) = \bU(\bX/\lambda,\tau)$ from (\ref{eq3_5}).
Let us apply the functional $\mathcal{A}$ to both sides in Eq. (\ref{eqC_1}). Using properties (\ref{eq_A_hom}) and (\ref{eq2_star}), we obtain
\be
    \label{eqC_2}
    \mathcal{A}[\bU_\lambda\left(\cdot,\tau\right)]
    = \dfrac{\alpha}{\lambda}
	\dfrac{A'_\ell(\bx_*'(t'),t')}{A_\ell(\bx_*(t),t)}.
\ee
Combining (\ref{eqC_1}) and (\ref{eqC_2}) yields the first hidden symmetry relation (\ref{eq3_4}). Similarly, using the relation $t' = \alpha t$ with (\ref{eq:newvariables}), (\ref{eq2_4}) and (\ref{eqC_2}), we derive
\be
    \begin{array}{rcl}
    d\tau' 
    & = &
	\displaystyle
    \dfrac{dt'}{T'_\ell(\bx_*'(t'),t')} 
    = \dfrac{A'_\ell(\bx_*'(t'),t') \,\alpha dt}{\ell}
    = \alpha\,\dfrac{A'_\ell(\bx_*'(t'),t')}{A_\ell(\bx_*(t),t)}
    \dfrac{A_\ell(\bx_*(t),t) dt}{\ell}
    \\[15pt] & = & \displaystyle 
    \lambda\left(
    \frac{\alpha}{\lambda}\dfrac{A'_\ell(\bx_*'(t'),t')}{A_\ell(\bx_*(t),t)} \right)
    \dfrac{dt}{T_\ell(\bx_*(t),t)}
    = \lambda \mathcal{A}[\bU_\lambda\left(\cdot,\tau\right)]
    d\tau.
    \end{array}
\ee
This yields the remaining hidden symmetry relation (\ref{eq3_5}). Thus, the fields $\bU(\bX,\tau)$ and $\bU'(\bX,\tau')$ are related by the hidden scaling symmetry.
Notice that the field $\bU'(\bX,\tau')$ does not depend on the temporal scaling factor $\alpha$.


\dataccess{The source code for the figures  is available at 
\url{https://github.com/sthalabard/HiddenSymmetryinNSIntermittency}. Version 1.0 used in this paper is archived at DOI:\\ 10.5281/zenodo.5036631.}

\aucontribute{ Both authors contributed equally to the manuscript. Both read and approved the manuscript.}

\competing{The author(s) declare that they have no competing interests.}

\funding{ST acknowledges support from  the Brazilian-French Network in Mathematics,  the
Programa de Capacitao Institucional of CNPq and  Fondation Louis D-Institut de France (project coordinated by M. Viana). AAM is supported by CNPq grants 303047/2018-6, 406431/2018-3.}

\ack{We thank S. Allende for graciously sharing the $512^3$ dataset.
We acknowledge H. Homann and C. Siewert for their essential help with the $4,096^3$ dataset, and 
thank J. Bec for related discussions.  }


\bibliographystyle{unsrt}
\bibliography{biblio}

\end{document}